\documentclass{article}

\usepackage{PRIMEarxiv}

\usepackage[utf8]{inputenc} 
\usepackage[T1]{fontenc}    
\usepackage{hyperref}       
\usepackage{url}            
\usepackage{booktabs} 
\usepackage{amsmath, amssymb, amsthm}
\usepackage{amsfonts}       
\usepackage{nicefrac}       
\usepackage{microtype}      
\usepackage{lipsum}
\usepackage{natbib}
\usepackage{hyperref}
\usepackage{fancyhdr}       
\usepackage{graphicx}       
\graphicspath{{media/}}     

\title{A Rank-Based Information Fusion Framework for
Comparing Clustered Multivariate Socioeconomic
Outcomes
\thanks{This manuscript has been accepted for publication at the 2026 IEEE International Conference on Fuzzy Systems (FUZZ-IEEE). The final published version will be available via IEEE Xplore.} 
}

\author{
  Dhrubajyoti Ghosh \\
  Kennesaw State University \\
  School of Data Science and Analytics \\
  Marietta, Georgia, USA\\
  \texttt{dghosh3@kennesaw.edu} \\
}

\begin{document}
\maketitle

\begin{abstract}
We propose a multivariate, distribution-free ranking framework for comparing clustered, correlated outcomes across groups, motivated by the evaluation of state-level policy environments using county-level socioeconomic data. Using pooled U.S. county data from 2019-2023, we study multiple dimensions of economic well-being, including poverty, income inequality, housing cost burden, medical care costs, and per capita income, observed at a finer spatial resolution than the policy itself. Rather than relying on parametric regression models, we employ a rank-based aggregation algorithm derived from the Longitudinal Rank-Sum Test (LRST), which treats clusters as independent units and aggregates information across outcomes using order statistics. This approach provides a robust, interpretable omnibus comparison that accommodates within-cluster dependence and high-dimensional outcome structure without distributional assumptions. Applied to the comparison of states with and without refundable Earned Income Tax Credit (EITC) policies, the method reveals systematic differences in the joint ranking of county-level outcomes, with results remaining stable under repeated random subsampling of counties and varying cluster sizes. 
While the empirical analysis is descriptive rather than causal, the study highlights the broader utility of rank-based, multi-criteria aggregation methods as computational intelligence tools for analyzing complex, clustered data in policy and social systems.
\end{abstract}


\section{Introduction}

State-level economic policies are frequently evaluated using aggregate indicators or parametric regression models that impose strong assumptions on functional form, distributional shape, and independence across observational units. In practice, many policies are implemented at the state level while economic outcomes are observed at finer geographic resolutions, such as counties, where outcomes are spatially clustered and highly correlated within states. Ignoring this structure can lead to misleading inference, particularly when multiple correlated outcomes are considered simultaneously \citep{conley1999gmm, anselin2022spatial}.
The Earned Income Tax Credit (EITC) is one of the largest income-support programs in the United States and has been widely studied in the applied economics literature. In addition to the federal program, many states have adopted refundable state-level EITCs that supplement federal benefits. Existing studies have examined the association between the EITC and labor supply, employment, earnings, and poverty outcomes, typically using regression-based identification strategies and focusing on a limited set of outcomes \citep{eissa1996labor,hoynes2018effective, neumark2011does}. While these studies provide important insights, they often rely on strong parametric assumptions and do not fully exploit the rich, high-dimensional information available at the county level.

This paper provides a complementary, data-driven assessment of state EITC policies by comparing county-level economic outcomes across states with and without a refundable EITC. Using comprehensive county-level data from PolicyMap covering the period 2019--2023, we examine multiple dimensions of economic well-being, including poverty rates, income inequality, housing cost burdens, medical care costs, and per capita income. Rather than estimating outcome-specific regression models, we adopt a multivariate, rank-based testing framework that aggregates information across outcomes and counties while explicitly accounting for within-state dependence.

Our empirical approach builds on the Longitudinal Rank-Sum Test (LRST), a nonparametric procedure designed for comparing groups in the presence of clustered and correlated multivariate outcomes \citep{xu2025novel, ghosh2025power, ghosh2025non}. Although originally developed for longitudinal and repeated-measures settings, the core idea of LRST extends naturally to cross-sectional contexts in which clusters, rather than individual observations, represent the primary independent units. In our setting, states serve as the independent units, while counties constitute dependent observations within each state. This framework allows us to test whether counties in EITC states tend to rank systematically better across multiple economic outcomes than counties in non-EITC states, without imposing distributional assumptions or specifying parametric models.
Applying this approach, we find consistent evidence that counties in states with a refundable EITC exhibit lower economic burden across a range of outcomes relative to counties in non-EITC states. While a single implementation yields marginal statistical significance, extensive robustness analyses based on repeated random subsampling of counties and sensitivity to the number of counties per state show that the findings are stable and highly robust. Overall, the results highlight the usefulness of multivariate, cluster-aware, nonparametric methods for policy evaluation using high-dimensional regional economic data.

The contribution of this work is the reinterpretation of rank-based U-statistics as an information fusion mechanism for clustered multivariate data. Specifically, we show how LRST can be adapted to settings where (i) clusters are the primary independent units, (ii) observations within clusters are dependent, and (iii) multiple correlated outcomes must be aggregated into a single decision statistic. This provides a distribution-free framework for multivariate policy comparison that avoids outcome-specific modeling and naturally accommodates within-cluster dependence.

\section{Data}

We use county-level economic data from PolicyMap covering the period 2019--2023. PolicyMap aggregates information from multiple administrative and survey-based sources, including the American Community Survey and other federal statistical agencies, and provides harmonized county-level indicators across the United States. Our analysis focuses on mainland U.S. counties and excludes Alaska, Hawaii, and U.S. territories.
We consider five economic outcomes that capture complementary dimensions of economic well-being: the poverty rate, the Gini index of income inequality, housing cost burden for renters, housing cost burden for homeowners, and per capita income. All outcomes are measured at the county level. To ensure a consistent interpretation across variables, outcomes are aligned so that higher values correspond to worse economic conditions. Specifically, per capita income is multiplied by minus one so that higher values indicate lower income.
These outcomes were selected to capture complementary dimensions of economic well-being, including income, inequality, cost burden, and access to resources, consistent with prior policy evaluation literature.
States are classified into two groups based on whether they operated a refundable state Earned Income Tax Credit during the study period. States with a refundable EITC are assigned to the treatment group, while all other states form the comparison group. Because economic outcomes are observed at the county level while policy is determined at the state level, counties within a state are treated as clustered observations.

\begin{figure}[ht]
\centering
\includegraphics[width=\linewidth]{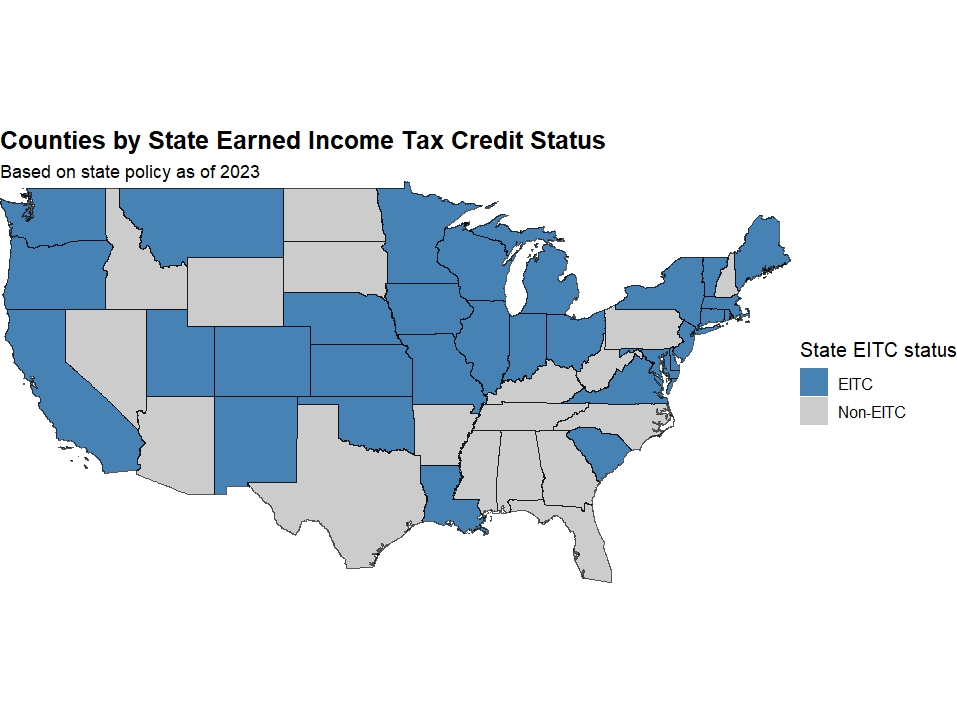}
\caption{Counties grouped by state Earned Income Tax Credit (EITC) status as of 2023. Blue indicates states with a refundable EITC; gray indicates states without a refundable EITC.}
\label{fig:eitc_map}
\end{figure}

Figure~\ref{fig:eitc_map} illustrates the geographic distribution of state Earned Income Tax Credit (EITC) policies across the continental United States as of 2023. States with a refundable EITC are shown in blue, while states without a refundable EITC are shown in gray. The figure highlights substantial spatial clustering in policy adoption, with EITC states concentrated in the Northeast, Midwest, and West Coast, and non-EITC states more prevalent in parts of the South and Mountain West. This clustering motivates our treatment of states as the independent units of analysis and counties as correlated observations nested within states in the subsequent empirical analysis.

\begin{figure}
    \centering
    \includegraphics[width=\linewidth]{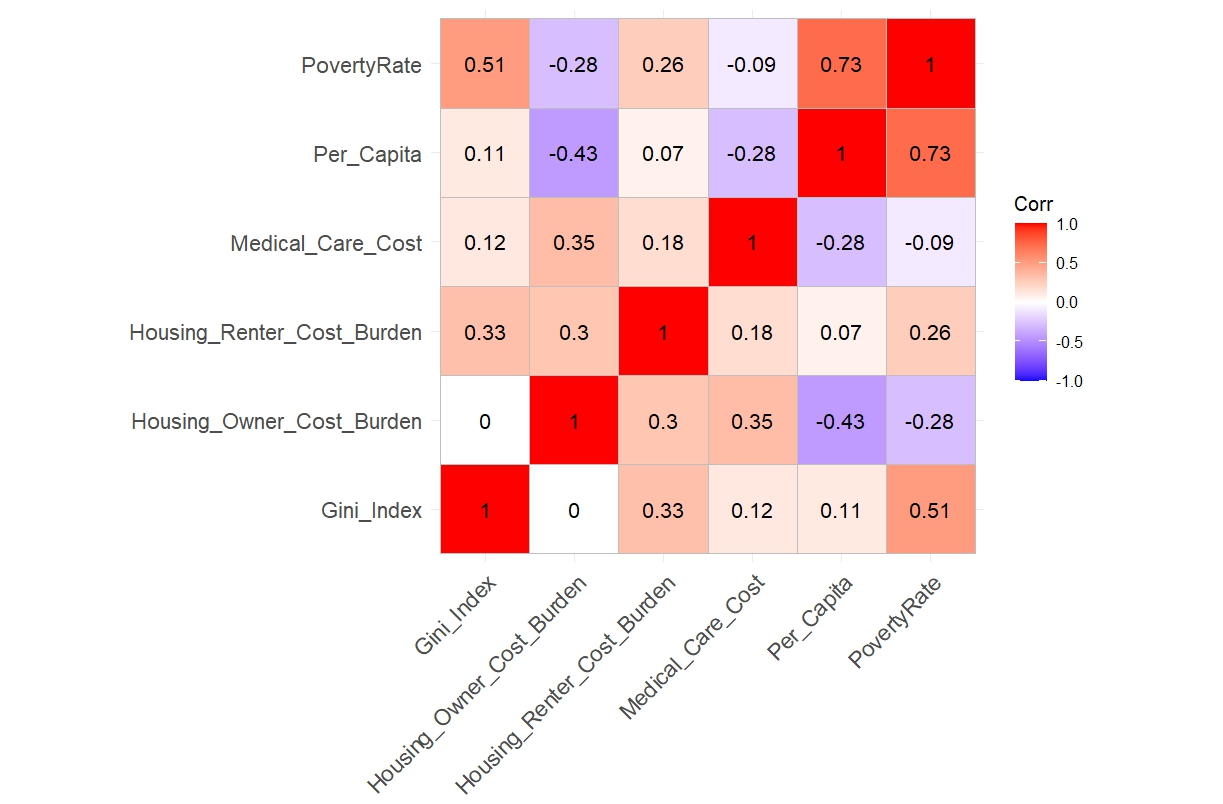}
    \caption{Correlation matrix of county-level economic outcomes. Variables are aligned so that higher values indicate worse conditions. The outcomes exhibit moderate but heterogeneous correlations, with a strong association between poverty and (negated) per capita income, and weaker or mixed relationships across other dimensions. This supports the need for multivariate aggregation.
    }
    \label{fig:corrPlot}
\end{figure}

The considered outcomes exhibit moderate but non-uniform correlations, indicating overlapping yet distinct dimensions of economic well-being (Figure~\ref{fig:corrPlot}). For example, poverty rate and (negated) per capita income are strongly correlated ($\rho \approx 0.73$), while income inequality shows moderate association with poverty ($\rho \approx 0.51$). In contrast, medical care costs and housing burdens display weaker and sometimes opposing relationships with other outcomes. This heterogeneous dependence structure motivates the use of a multivariate rank-based aggregation approach, as no single outcome fully captures the underlying economic conditions.

Missing county-level observations are imputed using the state-specific median for each outcome. This approach preserves within-state rank ordering and avoids imposing parametric distributional assumptions. 
Because the number of counties varies substantially across states, directly pooling all available counties would lead states with many counties to disproportionately influence the rank-based test statistic. To ensure that each state contributes equally to the comparison, we restrict attention to states with at least $C$ counties and draw a fixed-size subsample of exactly $C$ counties from each retained state. The same set of counties is used across all outcomes to preserve the multivariate dependence structure.
Counties are sampled randomly without replacement, and the baseline analysis uses $C = 50$. To assess sensitivity to this choice, we repeat the analysis for alternative values of $C$ and across multiple independent subsamples. This subsampling strategy ensures balanced state-level contributions while retaining the nonparametric and cluster-aware nature of the rank-based procedure.

\section{Empirical Approach}

Let $s = 1,\ldots,S$ index states and $i = 1,\ldots,n_s$ index counties within state $s$. For each county, we observe a $K$-dimensional vector of economic outcomes
\(
\mathbf{Y}_{si} = (Y_{si1}, \ldots, Y_{siK})^\top,
\)
and a binary policy indicator $Z_s \in \{0,1\}$, where $Z_s = 1$ denotes states with a refundable Earned Income Tax Credit (EITC).
Our objective is to test whether the county-level economic outcomes on average differ between EITC and non-EITC states while accounting for within-state dependence across counties and across outcomes. We adopt a multivariate, nonparametric rank-based testing framework based on the Longitudinal Rank-Sum Test (LRST) of \cite{xu2025novel}. Although LRST was originally developed for longitudinal data, its rank-based U-statistic structure extends naturally to clustered cross-sectional settings.

We treat states as independent experimental units and counties as dependent observations nested within states. For each outcome $k = 1,\ldots,K$, county-level observations are pooled across states and assigned mid-ranks $R_{sik}$, with larger values corresponding to more favorable economic conditions. Let
\(
\bar R_{sk} = \frac{1}{n_s} \sum_{i=1}^{n_s} R_{sik}
\)
denote the average rank for outcome $k$ within state $s$, and define the aggregated state-level rank
\(
\bar R_s = \frac{1}{K} \sum_{k=1}^K \bar R_{sk}.
\)

The global rank-sum statistic comparing EITC and non-EITC states is given by
\(
T = \frac{ \bar R^{(1)} - \bar R^{(0)} }
{ \sqrt{\widehat{\mathrm{Var}}(\bar R^{(1)} - \bar R^{(0)})} },
\)
where $\bar R^{(1)}$ and $\bar R^{(0)}$ denote the averages of $\bar R_s$ across EITC and non-EITC states, respectively.
The null hypothesis is
\(
H_0: \mathbf{Y}_{si} \stackrel{d}{=} \mathbf{Y}_{s'i'} \quad \text{for all } Z_s \neq Z_{s'},
\)
which states that 
the joint distribution of county-level outcomes is identical across the two policy groups. Under $H_0$, the statistic $T$ converges in distribution to a standard normal random variable. The variance estimator follows the cluster-adjusted U-statistic theory developed in \cite{xu2025novel} and accounts for dependence among counties within states as well as correlations across outcomes. The asymptotic validity of the test follows from standard cluster-level U-statistic theory under the regime where the number of clusters (states) tends to infinity while cluster sizes remain bounded or grow at a slower rate. Under these conditions, the test statistic converges to a normal distribution, with variance accounting for within-cluster dependence and cross-outcome correlations as established in the LRST framework.
The asymptotic validity of the test follows from standard cluster-level U-statistic theory under the regime where the number of clusters (states) tends to infinity while cluster sizes remain bounded or grow at a slower rate. Under these conditions, the test statistic converges to a normal distribution, with variance accounting for within-cluster dependence and cross-outcome correlations as established in the LRST framework \cite{xu2025novel, ghosh2025non}.
Rejection of $H_0$ indicates that counties in EITC states tend to rank systematically better across the collection of economic outcomes, providing a single omnibus test of policy-associated differences without requiring outcome-specific modeling or parametric assumptions. The LRST code is publicly available \cite{lrst_package}.

\section{Results}

\begin{figure}[ht]
\centering
\includegraphics[width=\linewidth]{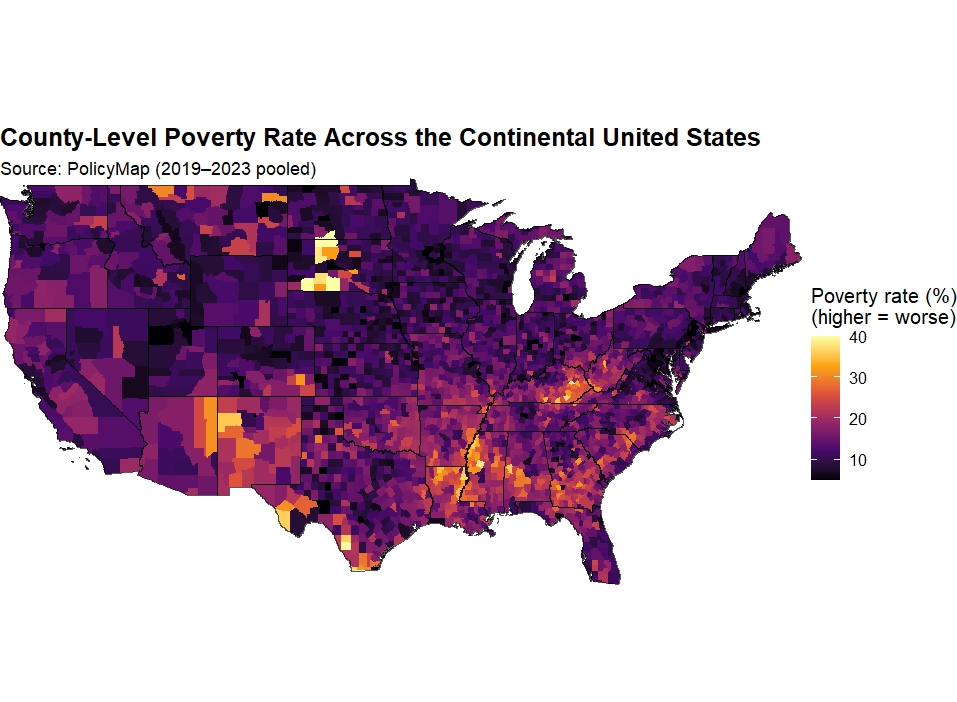}
\caption{County-level poverty rate across the continental United States. Lighter colors indicate higher poverty rates. Source: PolicyMap (2019--2023 pooled).}
\label{fig:poverty_map}
\end{figure}

We begin with exploratory analysis to characterize the spatial and distributional features of county-level economic outcomes. Figure~\ref{fig:poverty_map} presents a county-level choropleth map of poverty rates across the continental United States. The figure reveals pronounced spatial clustering, with persistently high-poverty regions concentrated in parts of the Southeast, Appalachia, and the Southwest, alongside relatively lower-poverty regions in the Midwest and Northeast. Similar geographic patterns are observed for other outcomes considered in this study. These spatial regularities highlight the importance of accounting for within-state dependence when conducting policy comparisons based on county-level data.

\begin{figure}[ht]
\centering
\includegraphics[width=0.9\linewidth]{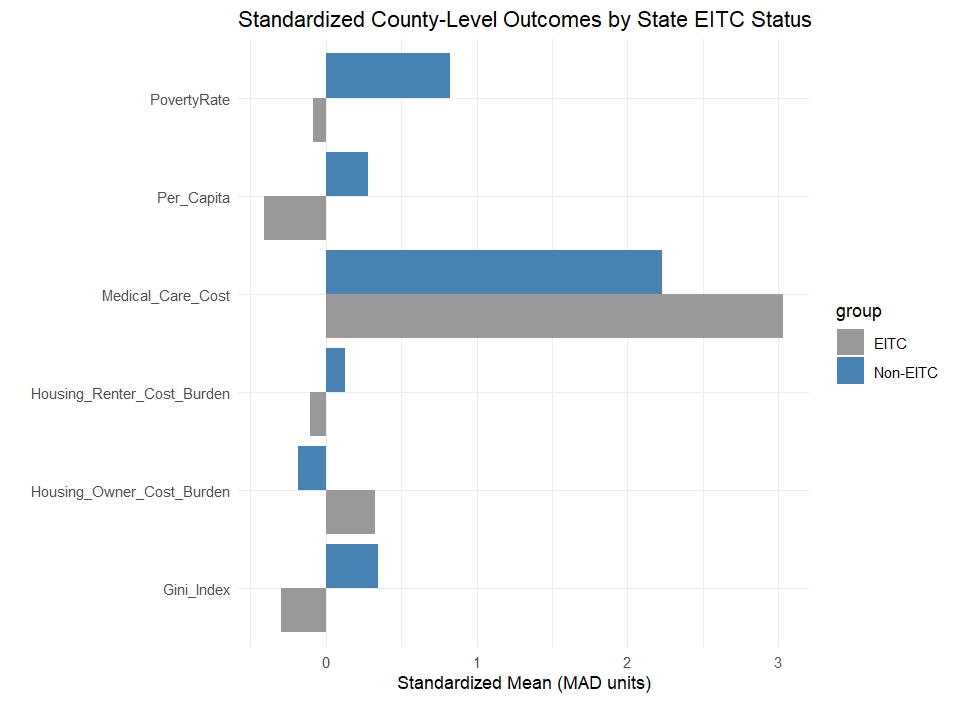}
\caption{Standardized mean differences in county-level economic outcomes between states with and without refundable EITC policies. Differences are expressed in units of the median absolute deviation. 
}
\label{fig:std_means}
\end{figure}

Figure~\ref{fig:std_means} summarizes standardized differences in county-level economic outcomes between EITC and non-EITC states. Differences are expressed in units of the median absolute deviation to facilitate comparison across outcomes measured on different scales. The figure shows that counties in EITC states tend to exhibit lower poverty rates, lower income inequality, and lower housing cost burdens, along with higher per capita income. Differences in medical care costs are also evident, although with greater dispersion.
While none of these standardized differences is large in isolation, the figure highlights a consistent directional pattern across outcomes. These modest but systematic shifts help explain why a multivariate rank-based test detects significant differences even when marginal distributions overlap substantially.


We next apply the Longitudinal Rank-Sum Test (LRST) to jointly compare county-level outcomes between EITC and non-EITC states. Using a baseline subsample of $C = 50$ counties per state, the LRST statistic is positive, indicating that counties in EITC states tend to rank better across the collection of economic outcomes than counties in non-EITC states. The corresponding p-value is marginally significant at conventional levels, providing initial evidence of systematic differences in county-level economic conditions associated with state EITC status.

To assess the stability of this finding, we repeat the analysis across 200 independent random subsamples of counties. The results are summarized in Table~\ref{tab:robustness}. The values of C were chosen to balance inclusion of a large number of states with sufficient within-state representation, while allowing sensitivity analysis across a reasonable range of cluster sizes. Across repetitions, the LRST statistic remains consistently positive with limited variability. For $C = 50$, the null hypothesis is rejected in 95\% of repetitions at the 5\% significance level and in all repetitions at the 10\% level. These findings indicate that the baseline result is not driven by a particular selection of counties, but instead reflects a stable pattern in the data.

\begin{table}[ht]
\centering
\caption{Robustness of LRST results to county subsampling}
\label{tab:robustness}
\begin{tabular}{lccccc}
\hline
$C$ & Mean $T$ & SD($T$) & Pr($p<0.10$) & Pr($p<0.05$) \\
\hline
30 & 10.4  & 1.09 & 1.00 & 0.755 \\
40 & 14.0  & 0.98 & 1.00 & 0.880 \\
50 & 17.6  & 0.95 & 1.00 & 0.945 \\
\hline
\end{tabular}
\end{table}

Table~\ref{tab:robustness} also reports sensitivity analyses with respect to the number of counties sampled per state. As $C$ increases from 30 to 50, the magnitude of the LRST statistic increases monotonically, and rejection frequencies at the 5\% level rise from 75.5\% to 94.5\%. Even for smaller subsample sizes, the test statistic remains positive in all repetitions, indicating directional consistency. Together, these results suggest that the evidence of systematically lower economic burden in EITC states is robust to both random county selection and the choice of subsample size.
Overall, the results provide consistent, multivariate evidence that counties in states with a refundable EITC tend to exhibit more favorable economic rankings across multiple dimensions, even though differences in any single outcome may be modest and masked by substantial cross-county heterogeneity.

\section{Discussion}

This study provides a multivariate, nonparametric comparison of county-level economic outcomes between U.S. states with and without a refundable Earned Income Tax Credit (EITC) \citep{nichols2015earned, hotz2001earned}. Our findings indicate that counties located in EITC states tend to rank systematically better across multiple dimensions of economic well-being, including poverty, housing cost burden, and income inequality. While differences in any single outcome are modest, they become more apparent when information is aggregated across outcomes and counties, underscoring the value of joint, multivariate assessment.

The exploratory analysis reveals pronounced spatial clustering and substantial within-state heterogeneity in county-level economic conditions, consistent with prior evidence documenting strong geographic disparities in income, poverty, and cost-of-living outcomes across the United States \citep{chetty2014land, autor2016china}. Such heterogeneity helps explain why marginal outcome distributions overlap considerably between EITC and non-EITC states and why univariate comparisons may lack power to detect systematic differences. In this setting, rank-based methods that aggregate information across correlated outcomes provide a natural complement to outcome-specific analyses.

Our results align with a large literature documenting associations between EITC policies and improved economic conditions for low- and moderate-income households, including reductions in poverty and income inequality \citep{eissa1996labor, leigh2006benefits, hoynes2018effective}. At the same time, the present analysis differs from much of the existing work by focusing on county-level outcomes and adopting a multivariate, cluster-aware perspective that does not rely on parametric modeling assumptions. Rather than estimating outcome-specific effects, the rank-based framework provides an omnibus assessment of whether economic conditions differ systematically across policy regimes.

Robustness analyses further support the stability of the findings. Repeated random subsampling of counties yields consistently positive test statistics with limited variability, and sensitivity analyses show that results strengthen as additional county-level information is incorporated. These patterns suggest that the observed differences are not driven by a small number of influential counties or particular sample constructions, but instead reflect persistent differences in the joint distribution of economic outcomes across states with differing EITC policies.

Several limitations merit discussion. First, the analysis is cross-sectional and descriptive, and the results should not be interpreted as causal effects of state EITC adoption. States that implement refundable EITCs may differ systematically from non-adopting states along institutional, demographic, or political dimensions not captured in the available data \citep{eissa2006behavioral}. Second, the analysis focuses on a selected set of economic indicators and does not incorporate labor market outcomes, migration responses, or demographic composition, which may also respond to state policy environments. Finally, while the rank-based approach is robust and flexible, it does not yield outcome-specific effect sizes, and thus complements rather than substitutes for regression-based causal analyses. Additionally, the framework can be extended to incorporate weighted aggregation across outcomes or clusters, depending on domain-specific priorities.

Despite these limitations, the findings illustrate the usefulness of multivariate, cluster-aware, nonparametric methods for policy evaluation using high-dimensional regional data. In settings where outcomes are spatially correlated and policy variation occurs at higher levels of aggregation, such methods can reveal systematic patterns that may be difficult to detect using conventional univariate approaches. Future work could extend this framework to other state-level policies or to longitudinal settings in which policy adoption varies over time.

As an alternative to rank-based approaches, one could also consider machine learning methods for aggregating multiple socioeconomic indicators, such as random forests \citep{breiman2001random}, gradient boosting \citep{natekin2013gradient}, or neural networks \citep{abdi1999neural}. These methods are capable of capturing complex nonlinear relationships and interactions across outcomes and may offer strong predictive performance \citep{ghosh2025ensemble, ghosh2026machine}. However, they typically require large training samples, rely on tuning parameters, and often lack interpretability in terms of a single, well-defined test statistic. Moreover, standard machine learning models are not designed to explicitly account for clustered dependence structures, such as counties nested within states, and do not provide distribution-free inference for group comparisons. In contrast, the proposed rank-based framework yields a transparent, assumption-light omnibus statistic that directly tests for systematic differences across groups while accommodating within-cluster dependence and multivariate outcome structure.

\bibliographystyle{unsrt}  
\bibliography{references}

@article{conley1999gmm,
  title={GMM estimation with cross sectional dependence},
  author={Conley, Timothy G},
  journal={Journal of econometrics},
  volume={92},
  number={1},
  pages={1--45},
  year={1999},
  publisher={Elsevier}
}

@article{anselin2022spatial,
  title={Spatial econometrics},
  author={Anselin, Luc},
  journal={Handbook of spatial analysis in the social sciences},
  pages={101--122},
  year={2022},
  publisher={Edward Elgar Publishing}
}

@article{eissa1996labor,
  title={Labor supply response to the earned income tax credit},
  author={Eissa, Nada and Liebman, Jeffrey B},
  journal={The quarterly journal of economics},
  volume={111},
  number={2},
  pages={605--637},
  year={1996},
  publisher={MIT Press}
}

@article{neumark2011does,
  title={Does a higher minimum wage enhance the effectiveness of the Earned Income Tax Credit?},
  author={Neumark, David and Wascher, William},
  journal={ILR Review},
  volume={64},
  number={4},
  pages={712--746},
  year={2011},
  publisher={SAGE Publications Sage CA: Los Angeles, CA}
}

@article{ghosh2025power,
  title={Power and sample size calculation for multivariate longitudinal trials using the longitudinal rank sum test},
  author={Ghosh, Dhrubajyoti and Xu, Xiaoming and Luo, Sheng and CPP Integrated Parkinson's Database},
  journal={Statistics in medicine},
  volume={44},
  number={20-22},
  pages={e70261},
  year={2025},
  publisher={Wiley Online Library}
}

@article{xu2025novel,
  title={A novel longitudinal rank-sum test for multiple primary endpoints in clinical trials: Applications to neurodegenerative disorders},
  author={Xu, Xiaoming and Ghosh, Dhrubajyoti and Luo, Sheng and CPP Integrated Parkinson’s Database},
  journal={Statistics in Biopharmaceutical Research},
  pages={1--11},
  year={2025},
  publisher={Taylor \& Francis}
}

@article{ghosh2025non,
  title={A non-parametric U-statistic testing approach for multi-arm clinical trials with multivariate longitudinal data},
  author={Ghosh, Dhrubajyoti and Luo, Sheng},
  journal={Journal of Multivariate Analysis},
  pages={105447},
  year={2025},
  publisher={Elsevier}
}

@article{chetty2014land,
  title={Where is the land of opportunity? The geography of intergenerational mobility in the United States},
  author={Chetty, Raj and Hendren, Nathaniel and Kline, Patrick and Saez, Emmanuel},
  journal={The quarterly journal of economics},
  volume={129},
  number={4},
  pages={1553--1623},
  year={2014},
  publisher={MIT Press}
}

@article{autor2016china,
  title={The China shock: Learning from labor-market adjustment to large changes in trade},
  author={Autor, David H and Dorn, David and Hanson, Gordon H},
  journal={Annual review of economics},
  volume={8},
  number={1},
  pages={205--240},
  year={2016},
  publisher={Annual Reviews}
}

@article{leigh2006benefits,
  title={Who benefits from the earned income tax credit? Incidence among recipients, coworkers and firms},
  author={Leigh, Andrew},
  year={2006},
  publisher={Walter de Gruyter}
}

@article{hoynes2018effective,
  title={Effective policy for reducing poverty and inequality?: the earned income tax credit and the distribution of income},
  author={Hoynes, Hilary W and Patel, Ankur J},
  journal={Journal of Human Resources},
  volume={53},
  number={4},
  pages={859--890},
  year={2018},
  publisher={University of Wisconsin Press}
}

@article{eissa2006behavioral,
  title={Behavioral responses to taxes: Lessons from the EITC and labor supply},
  author={Eissa, Nada and Hoynes, Hilary W},
  journal={Tax policy and the economy},
  volume={20},
  pages={73--110},
  year={2006},
  publisher={The MIT Press}
}

@article{ghosh2025ensemble,
  title={Ensemble survival analysis for preclinical cognitive decline prediction in Alzheimer's disease using longitudinal biomarkers},
  author={Ghosh, Dhrubajyoti and Pal, Samhita and Lutz, Michael and Luo, Sheng and Alzheimer’s Disease Neuroimaging Initiative},
  journal={Journal of Alzheimer’s Disease},
  year={2025},
  publisher={SAGE Publications Sage UK: London, England}
}

@article{breiman2001random,
  title={Random forests},
  author={Breiman, Leo},
  journal={Machine learning},
  volume={45},
  number={1},
  pages={5--32},
  year={2001},
  publisher={Springer}
}

@article{natekin2013gradient,
  title={Gradient boosting machines, a tutorial},
  author={Natekin, Alexey and Knoll, Alois},
  journal={Frontiers in neurorobotics},
  volume={7},
  pages={63623},
  year={2013},
  publisher={Frontiers}
}

@book{abdi1999neural,
  title={Neural networks},
  author={Abdi, Herv{\'e} and Valentin, Dominique and Edelman, Betty},
  number={124},
  year={1999},
  publisher={Sage}
}

@article{ghosh2026machine,
  title={Machine Learning Based Bot Detection on X With Temporal and Semantic Feature Integration},
  author={Ghosh, Dhrubajyoti and Boettcher, William and Johnston, Rob and Lahiri, Soumendra},
  journal={IEEE Transactions on Computational Social Systems},
  year={2026},
  publisher={IEEE}
}

@incollection{nichols2015earned,
  title={The earned income tax credit},
  author={Nichols, Austin and Rothstein, Jesse},
  booktitle={Economics of Means-Tested Transfer Programs in the United States, Volume 1},
  pages={137--218},
  year={2015},
  publisher={University of Chicago Press}
}

@misc{hotz2001earned,
  title={The earned income tax credit},
  author={Hotz, V Joseph and Scholz, John Karl},
  year={2001},
  publisher={National Bureau of Economic Research Cambridge, Mass., USA}
}

@software{lrst_package,
  author       = {Dhrubajyoti Ghosh},
  title        = {{LRST}: An {R} Package for Longitudinal Rank-Sum Test
                   on Multivariate Longitudinal Data
                  },
  month        = apr,
  year         = 2025,
  publisher    = {Zenodo},
  version      = {v0.1.0},
  doi          = {10.5281/zenodo.15182904},
  url          = {https://doi.org/10.5281/zenodo.15182904}
}

\end{document}